\documentclass[aps,pre,twocolumn,showpacs,superscriptaddress,groupedaddress]{revtex4}  % for review and submission
\usepackage{graphicx}  % needed for figures
\usepackage{dcolumn}   % needed for some tables
\usepackage{bm}        % for math
\usepackage{amssymb}   % for math
\usepackage[version=3]{mhchem}
\usepackage{color}

\usepackage[T1]{fontenc} 
\usepackage[utf8x]{inputenc}
\usepackage{amsmath}
\usepackage{natbib} 
\newcommand{\be}{\begin{equation}}
\newcommand{\ee}{\end{equation}}

\begin{document}

% The following information is for internal review, please remove them for submission
% \widetext
% \leftline{Version xx as of \today}
% \leftline{Primary authors: Joe E. Physics}
% \leftline{To be submitted to (PRL, PRD-RC, PRD, PLB; choose one.)}
% \leftline{Comment to {\tt d0-run2eb-nnn@fnal.gov} by xxx, yyy}
% \centerline{\em D\O\ INTERNAL DOCUMENT -- NOT FOR PUBLIC DISTRIBUTION}

% the following line is for submission, including submission to the arXiv!!
%\hspace{5.2in} \mbox{Fermilab-Pub-04/xxx-E}

 \title{Transcriptional leakage versus noise: A simple mechanism of conversion between
binary and graded response in autoregulated genes}

% \input author_list.tex       % D0 authors (remove the first 3 lines
%                              % of this file prior to submission, they
%                              % contain a time stamp for the authorlist)
%                              % (includes institutions and visitors)

% repeat the \author .. \affiliation  etc. as needed
% \email, \thanks, \homepage, \altaffiliation all apply to the current
% author. Explanatory text should go in the []'s, actual e-mail
% address or url should go in the {}'s for \email and \homepage.
% Please use the appropriate macro foreach each type of information

% \affiliation command applies to all authors since the last
% \affiliation command. The \affiliation command should follow the
% other information
% \affiliation can be followed by \email, \homepage, \thanks as well.

\author{Anna Ochab-Marcinek}
\email[]{ochab@ichf.edu.pl}
%\homepage[]{Your web page}
% \thanks{}
%\altaffiliation{}
\affiliation{Institute of Physical Chemistry, Polish Academy of Sciences, ul. Kasprzaka 44/52, 01-224 Warsaw, Poland}
%Collaboration name if desired (requires use of superscriptaddress
%option in \documentclass). \noaffiliation is required (may also be
%used with the \author command).
%\collaboration can be followed by \email, \homepage, \thanks as well.
%\collaboration{}
%\noaffiliation
\author{Marcin Tabaka}
\email[]{mtabaka@ichf.edu.pl}
%\email[]{Your e-mail address}
%\homepage[]{Your web page}
\thanks{The contribution of both authors, A.O.M. and M.T., to this work is equal. The manuscript has been published in Physical Review E, copyright APS, DOI: 10.1103/PhysRevE.91.012704, 
URL: \url{http://link.aps.org/doi/10.1103/PhysRevE.91.012704}}
%\altaffiliation{}
\affiliation{Institute of Physical Chemistry, Polish Academy of Sciences, ul. Kasprzaka 44/52, 01-224 Warsaw, Poland}

\date{\today}

\begin{abstract} 
We study the response of an autoregulated gene to a range of concentrations of signal molecules. We show that transcriptional leakage and noise due to translational bursting have the opposite effects. In a positively autoregulated gene, increasing the noise converts the response from graded to binary, while increasing the leakage converts the response from binary to graded. Our findings support the hypothesis that, being a common phenomenon, leaky expression may be a relatively easy way for evolutionary tuning of the type of gene response  without changing the type of regulation from positive to negative. 
\end{abstract}

\maketitle

% 87.18.Tt 	Noise in biological systems
% 87.16.Yc 	Regulatory genetic and chemical networks
% 87.18.Mp 	Signal transduction networks 
% 87.16.Xa 	Signal transduction and intracellular signaling

%\section{\label{sec:level1}First-level heading}
% sections are not used for PRL papers
\section{Introduction}
Leaky transcription (also called basal transcription) occurs when there is no tight control over the promoter and some level of transcription is maintained even when the promoter is in the \textit{off} state. To date, the role of transcriptional leakage has been underappreciated. Leaky expression is most often described as unfavorable from the point of view of an experimenter~\cite{Weber2007, Minaba2014,Guo2014}, while little is known about its evolutionary benefit for cells. An obvious fact is that some basal transcription is necessary to initiate the positive feedback~\cite{2005.avery.trends_microbiol}. Yanai \textit{et al.}~\cite{2006.yanai.trends_genet} note that, in general, the selection against ``unnecessary'' transcription is low and hypothesize that leakiness of the promoters may be evolutionarily neutral~ \cite{dekel05,shachrai10,szekely13}. On the other hand, Ingolia \textit{et al.}~\cite{2007.ingolia.curr_biol} put forward a hypothesis that, being a common phenomenon, leaky expression may be a relatively easy way 
for an evolutionary conversion of gene expression from binary to graded and vice versa. They qualitatively demonstrated in the experiments on yeast and in simulations that mutations in the  $P_{FUS1}$ promoter sequence entail changes in the basal level of expression of the autoregulated gene, which produces different expression patterns: unimodal or bimodal.

The effectors (signaling molecules) bind or cause phosphorylation of the transcription factors (TFs) thus changing the strength of gene repression or activation ~\cite{Rosenfeld05,2008.tabaka.jmb,Carey13}. We define the response to the increasing effector concentration as \textit{graded} when the stationary distribution of responses of individual cells is unimodal for any effector concentration. If bimodal distribution occurs for any range of those concentrations, then the response is \textit{binary}~\cite{1998.kringstein.pnas,2001.becskei.embo_j}. It became a common knowledge that positive autoregulation serves as a mechanism of differentiation of the cell population into phenotypically distinct groups~\cite{2007.alon.nat_rev_genet} (which may increase the chances of survival in a changing environment  through the bet-hedging strategy ~\cite{2009.fraser.mol_microbiol,Beaumont09}), while negative autoregulation may be preferred when a precise response is needed~\cite{Nevozhay2009,Little2013}.

Let us consider, however, an evolutionary adaptation from the conditions where binary response was favorable to the conditions where graded response is more preferred~\cite{2007.ingolia.curr_biol}. The evolutionary change of the nature of gene regulation from positive to negative may be more difficult than fine-tuning of the parameters of positive regulation, such as transcriptional leakage, e.g. due to point mutations in the promoter sequence~\cite{2007.ingolia.curr_biol,2010.mitrophanov.jmb}. 

Self-regulated genes often occur in two-component signaling systems (TCS) \cite{bijlsma2003making,2010.goulian.curr_opin_microbiol}. These systems respond  to external stimuli: Signal molecules bind to the membrane-bound receptors that phosphorylate the TFs, which enables the TFs to bind to the promoter of the target gene. TCS occur mostly in prokaryotes, and are less common in eukaryotes. The majority of TCS are positively autoregulated, but not all of them display bimodal expression \cite{2010.goulian.curr_opin_microbiol}. Known are the TCS with positive feedback and a significant basal transcription, e.g.: $hrpXY$ in \textit{E. amylovora} \cite{2000.wei.mpmi}, $CpxR$ in \textit{E. coli} \cite{2003.digiuseppe.j_bacteriol}, $VirG$ in \textit{Agrobacterium tumefaciens} \cite{Stachel1986,1987.yamamoto.mol_gen_genet}. 

Becskei \textit{et al.} \cite{2001.becskei.embo_j} first used the term ``conversion from graded to binary response'', but the conversion in their engineered $tetR$ gene circuit was obtained by artificially introducing a feedback loop into an otherwise open-loop system. Mitrophanov \textit{et al.} \cite{2010.mitrophanov.jmb} studied numerically and experimentally the positively autoregulated TCS $PhoP/PhoQ$ in \textit{Salmonella enterica} with promoter mutations resulting in different basal expression levels. They did not test a wide range of stimulus concentrations but only examined two cases of a low and high stimulus, and the binary response was not found. It is possible that these two stimulus levels represented the extreme cases (near-maximum and near-zero regulation) where the distributions are unimodal, and one should seek for the bimodal response by scanning the intermediate stimulus levels. Mitrophanov \textit{et al.} \cite{2010.mitrophanov.jmb} hypothesized that different evolutionary niches may favor higher or lower levels 
of basal expression and, consequently, different response levels.

\begin{figure}[t]
\begin{center}
\includegraphics[width=7cm]{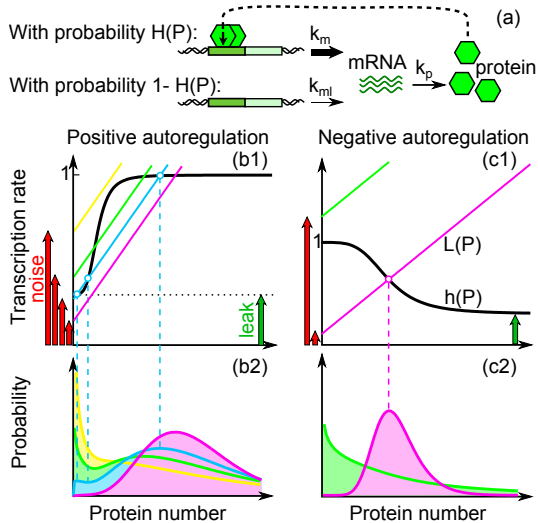}
\end{center}
\caption{\label{fig:schemat} {\small (Color online) The model of an autoregulated gene with leaky transcription [(a), positive regulation]. The intersections of the straight line $L(P)$ and the transfer function $h(P)$ (b1) and (c1) indicate the extrema of the protein distributions (b2) and (c2).}}
\end{figure}

We propose a simple quantitative model (Fig. \ref{fig:schemat}) of an autoregulated gene that allows one to calculate the conditions for conversion between binary and graded response without changing the type of feedback. We show the key role of intrinsic noise and leakage in this ``binary-graded'' conversion. 

\section{Model of an autoregulated gene}

We start from the kinetic scheme shown in Table~\ref{tab:kinetics} and make the simplifying assumptions, following Friedman \textit{et al.} ~\cite{2006.friedman.prl}: (i) mRNA is short-lived  compared to proteins.  This means that our simplified model may  be suitable for average prokaryotic genes but also for a subset of eukaryotic genes. The \textit{Escherichia coli} proteome shows insignificant degradation  \cite{koch1955protein} with a protein lifetime longer than the duration of a cell cycle but mRNA molecules are short-lived on the time scale of a cell cycle \cite{taniguchi2010quantifying}. In yeast, there is a certain percentage of genes that produce unstable mRNAs \cite{geisberg2014global}, and the mRNA stability in general depends on environmental conditions \cite{munchel2011dynamic}. At the same time, a substantial percentage of yeast proteins are long-lived \cite{belle2006quantification}. For example, the mean ratio $k_{dm}/k_{dp}$ in budding yeast is  $> 10$ (the median $\approx  3$) for the set of $ \sim 2000$ genes \cite{2008.shahrezaei.pnas}, showing that the model assumptions are widely met in this eukaryotic organism. (ii) The kinetics of TF binding and unbinding is fast enough to be compressed to the form of the Hill function
\be
H(P) = \frac{1}{1+cP^n},
\label{eq:O}
\ee
\begin{table}
% \begin{scriptsize}
\begin{tabular}{ c  c}
\multicolumn{2}{c}{\textbf{Transcription factor binding}:} \\
\textrm{Repressor} &  \textrm{Activator}\\
\ce{O + nP <=>[c] OP_n}& \ce{O + nP <=>[1/c] OP_n}\\
\multicolumn{2}{c}{\textbf{mRNA synthesis and degradation}:} \\  
 \textrm{Repressor} &  \textrm{Activator}\\
\ce{O ->[k_{\mathrm{m}}] M + O} &  \ce{O ->[k_{\mathrm{ml}}] M + O}\\
\ce{OP_n ->[k_{\mathrm{ml}}] M + OP_n} &  \ce{OP_n ->[k_{\mathrm{m}}] M + OP_n}\\
\multicolumn{2}{c}{\ce{M ->[k_{\mathrm{dm}}]   \varnothing}} \\
\multicolumn{2}{c}{\textbf{Transcription factor synthesis and degradation}:} \\
\multicolumn{2}{c}{\ce{M ->[k_{\mathrm{p}}] P + M}} \\
\multicolumn{2}{c}{\ce{P ->[k_{\mathrm{dp}}]   \varnothing}} \\
\end{tabular}
% \end{scriptsize}
\caption{\label{tab:kinetics}Model kinetics. $\mathrm{M}$, mRNA; P, protein; O, operator; n, number of TFs that bind cooperatively to the operator;  $c$, signal parameter; $k_m$, rate of mRNA synthesis from the  operator in the active state; $k_{ml}$, rate of mRNA synthesis from the operator in the inactive state (leakage); $k_{dm}$, rate of mRNA degradation ; $k_p$, rate of protein synthesis; $k_{dp}$, rate of protein degradation.}
\end{table}
where $P$ is the total number of TFs, $n$ is the cooperativity index, $n>0$ for repression and $n<0$ for activation. (iii) The cooperativity is very strong, such that only the $n$th power of $P$ is present in Eq. (\ref{eq:O}). The signal of a certain strength activates a certain fraction of TFs (due to phosphorylation, as in TCS, or due to binding of signal molecules). The signal parameter $c$ depends on the fractions of active and inactive TFs as well as on their binding and unbinding rates to the operator (see Appendix \ref{sec:sigparm}). The assumptions (i--iii) are necessary to make the model analytically tractable.

The operator effectively switches at a high frequency between the two states, \ce{O} and \ce{OP_n}, one of which is inactive and the other is active. The operator is active with the probability $H(P)$, and then the mRNA is synthesized at the rate $k_m$. Alternatively, the operator is inactive with the probability $1-H(P)$, and then the leaky transcription proceeds at the rate $k_{ml}<k_m$.  Therefore, at the steady state, $k_m H(P)  + k_{ml}(1-H(P)) = k_{dm}M$. We write the left-hand side divided by $k_m$ as 
\be h(P) = H(P)(1-\epsilon) + \epsilon,  \ee 
where $\epsilon = k_{ml} / k_m$. The function $h(P)$, which describes the deviation from the maximum possible transcription rate due to regulation and leakage, will be called the \textit{transfer function}~\cite{2010.ochab.pnas}. The deterministic equation for protein synthesis gives $k_p M = k_{dp} P$ at the steady state, where $k_p$ is the protein transcription rate and $k_{dp}$ is the protein degradation rate.  The deterministic stationary numbers of proteins can be then found by a geometric construction~\cite{2010.ochab.pnas},  as the points of intersections between the transfer function $h(P)$ and a straight line, $H(P)(1-\epsilon) + \epsilon = \frac{1}{\alpha \beta} P$, with $\alpha = k_m / k_{dp}$ and $\beta = k_p / k_{dm}$. If the straight line intersects the transfer function more than once, then the deterministic model is bistable. 
% 
%---------FIGURE ---------------------------------------------------------------
%---------------END OF FIGURE--------------------------------------------------
\section{Extrema of the protein distribution}
Based on the work of Friedman \textit{et al.}~\cite{2006.friedman.prl} we calculate the stochastic distribution of $P$ in the autoregulated gene with exponentially distributed translational bursts. We note that in ~\cite{2006.friedman.prl} the description of leakage in their transfer function, $\tilde{h}(P) = H(P) + \epsilon$, is only correct when $\epsilon \approx 0$, or if the parameters are reinterpreted  ($\tilde k_m \equiv k_m-k_{ml}$, $\tilde \epsilon \equiv k_{ml} / ( k_m-k_{ml})$, $\tilde \alpha \equiv  (k_m-k_{ml}) / k_{dp}$; see Appendix \ref{sec:reint}), since otherwise the probabilities of the active and inactive states of the operator would not sum up to 1. Our description is the most natural, as the parameters are simply the reaction rates that follow directly from the kinetics (Table \ref{tab:kinetics}). Therefore, the protein distribution in our model differs from that in~\cite{2006.friedman.prl}:
\begin{align}\label{eq:friedman_rozklad}
p(P)& = A P^{-1} e^{-P/\beta} e^{\alpha \int dP \ h(P)/P}\\ \nonumber
&= A P^{\alpha - 1} e^{-P/\beta}  H(P)^{\alpha(1-\epsilon) / n}. 
\end{align}
Here $\alpha=k_m/k_{dp}$, $\beta = k_p/k_{dm}$ is interpreted as the mean burst size, and $A$ is a normalization constant.  Note also that $\alpha$ should be interpreted as the \textit{maximum} mean frequency of translational bursts that can be achieved by the system in the theoretical limit of $c=0$ (see Appendix \ref{sec:mean_freq}). In a certain range of the regulation strength $c$, the distribution $p(P)$ can be bimodal. The conditions for bimodality can be determined by calculation of the extrema of the distribution:  $\frac{dp(P)}{dP}=0$ ~\cite{2010.ochab.pnas,Mackey2010,2012.aquino.pre}. In the present stochastic model, the extrema of $p(P)$ are again given by  the points of intersections between the transfer function $h(P)$ and a straight line (which we will denote by $L(P)$):
\be\label{eq:stoch}
H(P)(1-\epsilon) + \epsilon = \frac{1}{\alpha \beta} P + \frac{1}{\alpha}
\ee
The geometric construction is almost the same as in the deterministic case but there appears the noise term $1/\alpha$. It shifts the positions of the extrema with respect to the deterministic stationary states. Therefore, the noise may induce a bifurcation in the parameter range in which the deterministic model does not predict bistability \cite{horsthemke_lefever.book}. Interestingly, this  noise-induced shift depends on the maximum burst frequency $\alpha$ only, and not on the burst size $\beta$. In the limit of infinitely frequent bursting, the stochastic term $1/\alpha$ disappears and the extrema of the protein distribution overlap with the deterministic stationary states. Therefore, we will use $1/\alpha$ as a measure of the minimum noise that can be achieved by the system at the theoretical limit of $c=0$.

\section{Binary and graded response to a signal}\label{sec:main_results}
Using the geometric construction (\ref{eq:stoch}), we examine how the protein distributions behave when the regulation strength $c$ is varied. The change in $c$ corresponds to the change in the effector concentration which controls the strength of gene regulation. To obtain precise regulation, the gene response should be graded for the whole possible range of $c$. The following conclusions follow from the geometric construction (see Appendix \ref{sec:detail} for a detailed analysis): Transcriptional leakage narrows the range of regulation. Negative feedback allows for a graded response only, because at most one intersection of $L(P)$  and $h(P)$ is possible due to their different monotonicity. When the feedback is positive, binary response is possible. For $ \epsilon < 1/\alpha<1$, the response is always binary (also for $n=-1$, consistently with the results of \cite{2006.friedman.prl}, the case not predicted by the deterministic model).

Our central result is that transcriptional leakage counteracts the stochastic effect of bursting, by converting binary response into graded response. This is because $\epsilon>0$ shifts the base of the transfer function upwards, in such a way that, if the leakage is sufficiently large, it may enable only one intersection of $h(P)$ and $L(P)$.
%  \ref{fig:n1binary}
%---------FIGURE ---------------------------------------------------------------
\begin{figure}[t]
\begin{center}
\includegraphics[width=8.5cm]{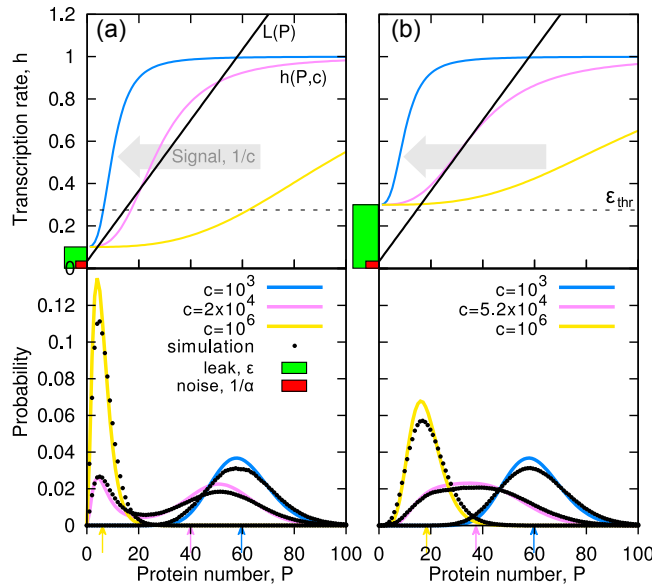}
\end{center}
\caption{\label{fig:nbinary} {\small (Color online) Transcriptional leakage converts binary response into graded response and counteracts the effect of noise. Binary response occurs for the leakage below the threshold  $\epsilon_{thr}$, when $1/ \alpha <1$ (dashed line). Otherwise, graded response occurs. (Dots) Mesoscopic simulations of the gene regulation kinetics using the Gillespie algorithm~\cite{gibson2000efficient}). (a) $\epsilon=0.1$. (b) $\epsilon=0.3$. Parameters are as follows: $n=-3$, $\alpha=30$, $\beta=2$, $\epsilon_{thr}=0.275$. (Arrows) Mean values of the distributions.}}
\end{figure}
%---------------END OF FIGURE--------------------------------------------------
Below, we calculate the condition for the graded response in the general case of $n<0$, when $1/ \alpha <1$. We note that for given  $\alpha$ and $\beta$ there exists a value $c=c^*$ for which $L(P)$ intersects $h(P,c^*)$ in its inflection point. The inflection point is the point in which $h(P)$ has the greatest slope, and this slope increases monotonically as $c$ decreases. Therefore, if the slope of $L(P)$ intersecting  $h(P,c^*)$ in its inflection point is greater than the slope of $h(P,c^*)$, then $L(P)$ will intersect  $h(P,c)$ only once for any $c$. We write this condition as:
\begin{align} \label{eq:cond1}
 h(P_p,c^*)=\frac{1}{\alpha \beta} P_p + \frac{1}{\alpha}& \  \mathrm{and} & \frac{d}{d P_p} h(P_p,c^*)<\frac{1}{\alpha \beta},
\end{align}
where $P_p = \left[ (n-1)/\left( c^*(n+1) \right)\right] ^{1/n}$ is the value of $P$ in the inflection point. Knowing that $H(P_p) = (n+1)/(2n)$ for any $c$, and  $H'(P_p) = - (n^2-1)/(4nP_p)$, we get:
\be \label{eq:cond}
 \epsilon >  \frac{1}{\alpha}\frac{(- 4n)}{(n-1)^2}+ \frac{ (n+1)^2}{(n-1)^2}\equiv \epsilon_{thr} 
\ee
The above condition says that if the transcriptional leakage $\epsilon$ is greater than the threshold $ \epsilon_{thr} $, then the positively autoregulated gene will produce a graded response. For the leakage below that threshold, the response will be binary (Fig. \ref{fig:nbinary}). The condition (\ref{eq:cond}) depends only on the cooperativity $n$ and the maximum burst frequency $\alpha$, but not on the burst size $\beta$. Since $n<0$ for activation, both the noise and the cooperativity increase $ \epsilon_{thr} $, which makes the graded response more difficult to obtain. This finding is consistent with the experimental observation of Ingolia \textit{et al.}~\cite{2007.ingolia.curr_biol} that bimodal distributions are found for low basal expression and high induced expression (which corresponds to a steeper transfer function).

\section{Conclusions}
The above calculation based on a geometric construction reveals the constructive role of transcriptional leakage in gene regulation: The leakage in a positively autoregulated gene acts against the translational noise as a factor that controls the conversion between binary and graded response. While increasing the noise induces binary response, increasing the leakage recovers graded response. However, this conversion is obtained at the cost of narrowing the range of regulation. Therefore, the leakage can be disadvantageous in the case of negative autoregulation (because the response is anyway graded), but it can be beneficial in the case of positive autoregulation, when it is needed to prevent the binary response. Leakage strength can be tuned by single mutations in the promoter~\cite{2007.ingolia.curr_biol}, whereas  keeping the same direction of the response after the positive to negative feedback conversion would also require the reversal of the signal effect on the TF (if a high concentration of signal 
molecules strengthened the binding of the activator to the promoter, now it should cause a weaker binding of the repressor). The  conversion of the feedback type is thus a much less probable evolutionary scenario because it would  require multiple mutations (within the TF's effector binding site and its DNA-binding domain) while simultaneously keeping the TF function. Our findings may  therefore provide a quantitative support for the experimentally based hypothesis~\cite{2007.ingolia.curr_biol,2010.mitrophanov.jmb} that, being a common phenomenon, leaky expression can be an easier way of adaptation of the gene response type to different evolutionary niches than the change of the feedback type from positive to negative. 

\begin{acknowledgments}
A.O.M. was supported by the Ministry of Science and Higher Education grant no. 0501/IP1/2013/72 (Iuventus Plus).
\end{acknowledgments}

\appendix
\section{Signal parameter $c$}\label{sec:sigparm}
Below, we show how the coefficient $c$ in the Hill function $H(P)$ contains the information about the signal strength. For simplicity, we take the example of $n=1$. $P = P_a + P_i$ denotes the total number of transcription factors (TFs), both active and inactive. $k_{on}^a$, $k_{off}^a$ denote the binding and unbinding rates of the active TF to the operator. $k_{on}^i$, $k_{off}^i$ are the binding and unbinding rates for the inactive TF. Then,
\begin{align} \label{eq:exn1}
H(P) = &\frac{1}{1 + \frac{k_{on}^a}{k_{off}^a}P_a + \frac{k_{on}^i}{k_{off}^i}P_i}\\ \nonumber
        = &\frac{1}{1 + P \left[\frac{k_{on}^a}{k_{off}^a}f_a + \frac{k_{on}^i}{k_{off}^i}(1-f_a)\right]}\\ \nonumber
         = &\frac{1}{1 + cP}.
\end{align}
$f_a$ denotes the active fraction of all TFs, e.g. the fraction of phosphorylated TFs in the case of a two-component system (assuming that phosphorylation and dephosphorylation rates are such that this fraction remains constant on the time scales of other reactions). In the case of binding of a signal molecule (effector) \ce{E}, $f_a = g(E)$ is a Hill function describing the effector binding to the TF (under the assumption that $E \gg P$ and that the effector-TF binding and unbinding rates are much faster than the time scales of other reactions in the system).

For $n>1$, $H(P)$ is constructed in an analogous way under the assumption of very strong cooperativity. Although mixed terms may appear [e.g. for $n=2$, $P_aP_i( k_{on,1}^a k_{on,2}^i)/(k_{off,1}^ak_{off,2}^i)$ etc., where the number in the subscript denotes binding of the first or the second TF], they can still be written in the form of $P^n \cdot const$. And therefore, the coefficient $c$ will still contain the particular binding/unbinding rates and the details of the TF-effector interactions.

\section{Limiting cases}\label{sec:limits}
 Note that $c=c(f_a)$ varies between finite values as the fraction $f_a$ varies from 0 to 1. In the case of $n=1$, as in the example (\ref{eq:exn1}), $c(f_a=0)=k_{on}^i / k_{off}^i$ and $c(f_a=1)=k_{on}^a/ k_{off}^a$. This means that $H(P)=0$ and $H(P)=1$ are the theoretical limits for an extremely strong or weak signal when the binding of active TFs is infinitely fast and the binding of inactive TFs is infinitely slow compared to unbinding. These two limits correspond to non-regulated genes \cite{2006.friedman.prl,2008.shahrezaei.pnas}, i.e. Gamma distributions with the means $\alpha \beta$ and $\alpha \beta \epsilon$.
%---------FIGURE ---------------------------------------------------------------
\begin{figure}[t!]
\begin{center}
\includegraphics[width=4cm]{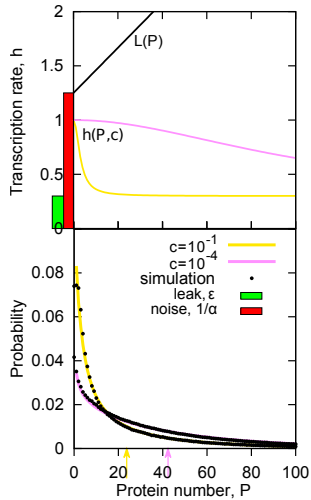}
\end{center}
\caption{\label{fig:alpha_less_one} {\small (Color online) When bursts are rare, $\alpha<1$, then the maximum of the protein distribution $p(P)$ is always at $P_{max}=0$ independently of the leakage rate $\epsilon$ and the regulation strength $c$. This is because there is no intersections between the straight line $L(P)$  and the transfer function $h(P)$. Parameter values are as follows:
$n=2$, $\alpha=0.8$, $\beta=60$, $\epsilon=0.3$. Arrows mark the mean values of the shown distributions.}}
\end{figure}
%---------------END OF FIGURE--------------------------------------------------
\section{Reinterpretation of the parameters of the formula used in~\cite{2006.friedman.prl} }\label{sec:reint}
The formula proposed by Friedman \textit{et al.}~\cite{2006.friedman.prl} to describe the protein distribution produced by an autoregulated gene,
\be\label{eq:friedman_rozklad_orig}
p(P) = A P^{\tilde \alpha (1+ \tilde \epsilon)  - 1} e^{-P/\beta}  H(P)^{\tilde \alpha / n},
\ee
can be made formally equivalent to our formula (\ref{eq:friedman_rozklad}), if the parameters of (\ref{eq:friedman_rozklad_orig}) are reinterpreted.

In our model, the total transcription rate is:
\begin{align} \label{eq:1}
k_m H(P)  + k_{ml}(1-H(P)) =&\\ \nonumber
=& H(P)(k_m - k_{ml}) + k_{ml}\\ \nonumber
\equiv & k_m [H(P)(1-\epsilon) + \epsilon] \\ \nonumber
\equiv &  k_m h(P)
\end{align}
where $k_m$ is the transcription rate in the case when the operator is in the active state. This is the maximum possible transcription rate, which can be achieved if the operator is \textit{on} all the time. $k_{ml}$ is the transcription rate in the case when the operator is in the inactive state (leaky transcription). $\epsilon = k_{ml} / k_m$  measures the ratio of the transcription rate  in the inactive state to the transcription rate in the active state. The transfer function $h(P)$ measures the deviation from the maximum possible transcription rate due to regulation and leakage.

Friedman \textit{et al.} use a different transfer function: $\tilde h (P) = H(P) + \tilde \epsilon$. The total transcription rate is then 
\be \label{eq:2}
 \tilde k_m [H(P) + \tilde \epsilon] \equiv \tilde k_m \tilde h(P).
\ee
In order to make it equivalent to (\ref{eq:1}), one must assume $\tilde k_m = k_m - k_{ml}$ and $\tilde \epsilon = k_{ml} / \tilde k_m$. With this interpretation of the parameters, the formulas (\ref{eq:friedman_rozklad_orig}) and (\ref{eq:friedman_rozklad}) for the protein distributions will be equivalent, provided that $\tilde \alpha = \tilde k_m / k_{dp}$, whereas in our model $\alpha = k_m / k_{dp}$.

It should be noted that  $\tilde k_m$ \textit{is not} the transcription rate $k_m$ shown in the kinetic scheme (Table I in the main manuscript). Therefore, our notation (without tilde) is more natural because it follows directly from the kinetics.

At this point, we note that neither the interpretation of $\tilde \alpha$ nor $\alpha$ as ``the mean number of bursts per cell cycle'' (as in~\cite{2006.friedman.prl}) is fully accurate. We clarify the interpretation of $\alpha$ in the next section.
%---------FIGURE ---------------------------------------------------------------
\begin{figure}[t]
\begin{center}
\includegraphics[width=8cm]{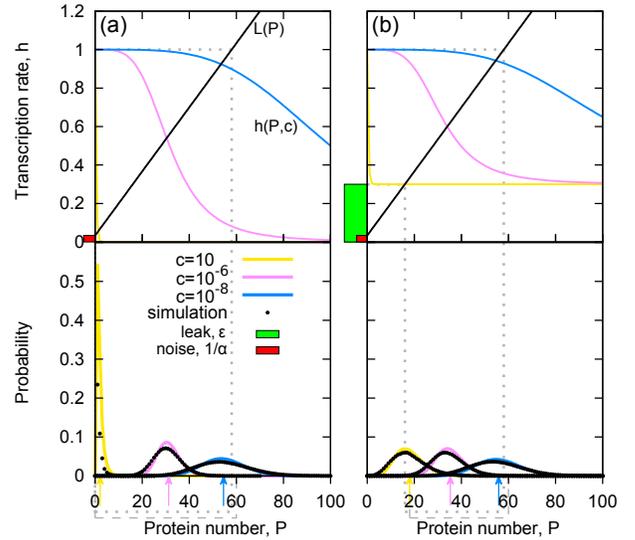}
\end{center}
\caption{\label{fig:narrow} {\small (Color online) Transcriptional leakage $\epsilon>1/\alpha$ narrows the range of maxima and mean values of protein distributions. (a) No leakage, $\epsilon=0$. (b) Leakage $\epsilon=0.3$. (Dotted lines) Range of maxima. (Dashed lines) Range of mean values. (Arrows) Mean values of the shown distributions. Parameters are as follows: $n=4$, $\alpha=30$, $\beta=2$. }}
\end{figure}
%---------------END OF FIGURE--------------------------------------------------
\section{Mean burst frequency}\label{sec:mean_freq}
Our model assumes that the promoter effectively switches between the states \ce{O} and \ce{OP_n}, and the switching is very fast. Transcription occurs as one of the two alternative processes (here shown for the case of repression): 
\be
\begin{cases}
\ce{O ->[k_{\mathrm{m}}] M + O} \ \ \mathrm{(with \ the \ probability \ } H(P)\mathrm{)}\\ 
\ce{OP_n ->[k_{\mathrm{ml}}] M + OP_n} \ \ \mathrm{(with \ the \ probability \ } 1-H(P)\mathrm{)}
\end{cases}
\ee

The mean burst frequency $\nu$ for an autoregulated gene in a stationary state yields then
%---------FIGURE ---------------------------------------------------------------
\begin{figure}[t!]
\begin{center}
\includegraphics[width=8cm]{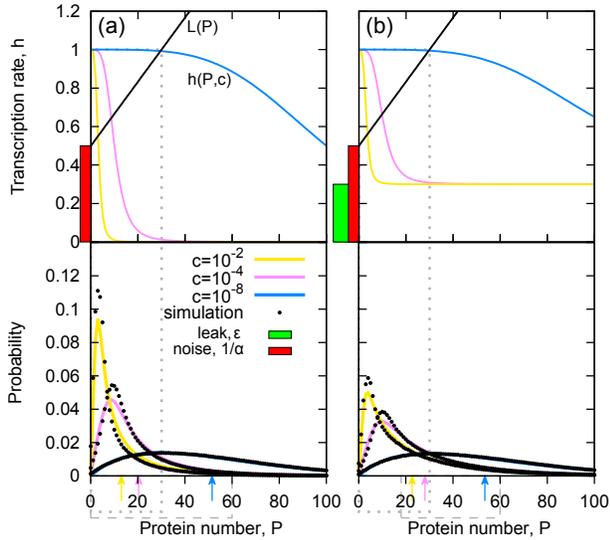}
\end{center}
\caption{\label{fig:nonarrow} {\small (Color online) When the transcriptional noise is sufficiently high, $1/\alpha > \epsilon$, the leakage does not narrow the range of maxima (dotted lines). However, the range of mean values (dashed lines) is narrower for nonzero leakage. Arrows: mean values of the shown distributions.  (a) No leakage, $\epsilon=0$. (b) Leakage $\epsilon=0.3$. Parameters: $n=4$, $\alpha=2$, $\beta=30$.}}
\end{figure}
%---------------END OF FIGURE--------------------------------------------------
%---------FIGURE ---------------------------------------------------------------
\begin{figure}[t]
\begin{center}
\includegraphics[width=8cm]{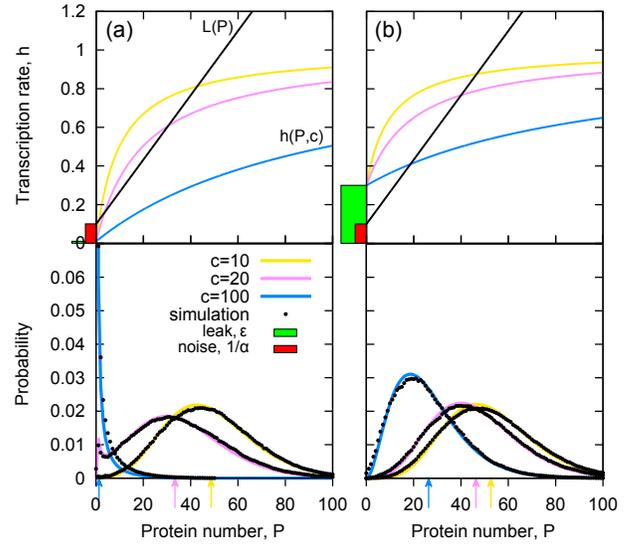}
\end{center}
\caption{\label{fig:n1binary} {\small (Color online) (a) On the contrary to the deterministic model, binary response is possible in the stochastic model due to transcriptional bursting, even when there is just one transcription factor binding site. Here, only a very small leakage is present ($\epsilon=0.01$). (b) Transcriptional leakage ($\epsilon=0.3$) counteracts the stochastic effect of bursting by converting binary response into graded response. Parameters: $n=-1$, $\alpha=10$, $\beta=6$. Arrows: mean values of the shown distributions.}}
\end{figure}
%---------------END OF FIGURE--------------------------------------------------
\begin{align} \label{eq:nu}
\nu& = \frac{1}{k_{dp}} \left( k_m  \left\langle H(P)\right\rangle_{p(P)} +  k_{ml} \left\langle 1-H(P) \right\rangle_{p(P)} \right)\\
&=\left\langle H(P)\right\rangle_{p(P)}\frac{k_m-k_{ml}}{k_{dp}} + \frac{k_{ml}}{k_{dp}},
\end{align}
where $p(P)$ is the protein number distribution (\ref{eq:friedman_rozklad}), and the average
\be\label{eq:havg}
\left\langle H(P)\right\rangle_{p(P)} = \int_0^\infty H(P) p(P) dP
\ee
is used because of the assumption of a rapid switching of the promoter state. The value of (\ref{eq:havg}) for a given $c$  lies between 0 and 1. Consequently, the mean burst frequency $\nu$ depends on the signal level $c$ and it lies between $k_m/k_{dp}$ and $k_{ml}/k_{dp}$.

We therefore interpret $\alpha = k_m/k_{dp}$  as the \textit{maximum}  mean frequency of translational bursts that can be achieved by the system. This occurs in the theoretical limit of $c=0$, i.e. when $H(P)=1$ (see Appendix \ref{sec:limits} above). The parameter $1/\alpha$ is then interpreted as a measure of the \textit{minimum} noise that can be achieved by the system.

\section{Detailed analysis of the geometric construction}\label{sec:detail}
(1) When the bursts are rare, $\alpha<1$, then, for both negative and positive feedback, the maximum of the protein distribution is always at $P_{max}=0$ independently of the leakage rate $\epsilon$ and the regulation strength $c$.  In the deterministic model there always exist values of $c$ at which a peak occurs at $P_{max}\neq0$. But the presence of the stochastic term $1/ \alpha>1$ makes it impossible for the straight line $L(P)$ to intersect $h(P)$ for any value of $c$ (Fig. \ref{fig:alpha_less_one}). Varying $c$ makes the distributions only narrower or wider, which varies the mean protein number in the range ($\alpha \beta \epsilon$, $\alpha \beta$).

(2) Negative feedback allows for graded response only. When the bursts are frequent, $\alpha>1$, there always is one intersection of $L(P)$  and $h(P)$ because of their different monotonicity. Transcriptional leakage narrows the range of regulation (Fig. \ref{fig:narrow}): the maximum of $p(P)$ has the range $P_{max} \in [ \max(0,\epsilon \alpha \beta - \beta); \alpha \beta - \beta]$, and the mean has the range $\langle P \rangle  \in [\alpha \beta \epsilon; \alpha \beta]$. However, when the noise term $1/\alpha > \epsilon$, then the range of $P_{max}$ does not depend on $\epsilon$ (Fig. \ref{fig:nonarrow}). Therefore, a sufficiently strong noise (long mean time between random bursts) counteracts the negative effect of leakage on the range of maxima at the cost of wider distributions (at a given $\alpha\beta$), but not on the range of mean values.

(3) When the feedback is positive and  $ \epsilon < 1/\alpha < 1 $, binary response is always present because there always is a range of the signal levels in which the protein distribution is bimodal [Fig. \ref{fig:n1binary}A]. Note that, consistently with the results of \cite{2006.friedman.prl}, the bursting term $1/\alpha$ makes it possible to obtain binary response also for $n=-1$ [Fig. \ref{fig:n1binary}A], the case not predicted by the deterministic model.

(4) Transcriptional leakage counteracts the stochastic effect of bursting, by converting binary response into graded response. This is because $\epsilon>0$ shifts the base of the transfer function upwards, in such a way that, if the leakage is sufficiently large, it may enable only one intersection of $h(P)$ and $L(P)$. When the regulation is positive with $n=-1$, it suffices that the transcriptional leakage $\epsilon> 1/\alpha$ to obtain graded response [Fig. \ref{fig:n1binary}B]. This condition is insufficient when $n<-1$. In the main text (Sec. \ref{sec:main_results}), we calculated the condition for graded response in the general case of $n<0$, when $1/\alpha < 1$.
\section{Simulation parameters}\label{sec:simul}
Reaction rate constants used in the mesoscopic simulations (Gillespie algorithm~\cite{gibson2000efficient}) of the gene regulation kinetics have been shown in Table \ref{tab:kinetic_parameter}. We model the signal parameter $c$ as the ratio of effective rate constants $(k_{on,1}...k_{on,n})/(k_{off,1}...k_{off,n})$ for TF association and dissociation to the binding sites $1 ..n$ on the operator. The effective rates mimic the influence of the effectors or phosphorylation on the ratio of active TFs.
\begin{table*}[t!]
\footnotesize{
   \begin{tabular}{c|ccccc|c|c}
   Figure &  $k_{\mathrm{m}}$   & $k_{\mathrm{ml}}$    & $k_{\mathrm{dm}}$   &  $k_{\mathrm{p}}$   & $k_{\mathrm{dp}}$ &  $k_{\mathrm{on}}$   & $k_{\mathrm{off}}$ \\\hline
    2.A   &  $3\cdot10^{-3}$   & $3\cdot10^{-4}$      & $10^{-3}$         &   $2\cdot10^{-3}$    & $10^{-4}$                           & $c=10^{3}$: \{$10^{-2}, 5\cdot10^{-1}$,1\}; $c=2\cdot10^{4}$: $\{10^{-3}, 5\cdot10^{-2}, 5\}$ & $\{1,5,1\}$\\
             &                              &                                &                         &                                &                                           &   $c=10^{6}$: $\{10^{-4}, 5\cdot10^{-2}, 1\}$                     &                                                     \\\cline{7-8}
    2.B   &  $3\cdot10^{-3}$   & $9\cdot10^{-4}$      & $10^{-3}$         &   $2\cdot10^{-3}$    & $10^{-4}$                           & $c=10^{3}$: \{$10^{-2}, 5\cdot10^{-1}$,1\}; $c=5.2\cdot10^{4}$: $\{10^{-3}, 5\cdot10^{-2}, 10\}$ & $\{1,5,1\}$; \\
             &                              &                                &                         &                                &                                           &   $c=10^{6}$: $\{10^{-4}, 5\cdot10^{-2}, 1\}$                     &                        $c=5.2\cdot10^{4}$:  $\{1,5,5.2\}$    \\\cline{7-8}
   S1       &  $0.8\cdot10^{-4}$   & $2.4\cdot10^{-5}$      & $10^{-3}$         &   $6\cdot10^{-2}$    & $10^{-4}$                           & $c=10^{-1}$: \{$10^{-2}, 10^{-1}$\}; $c=10^{-4}$: $\{10^{-2}, 10^{-1}\}$ & $\{10^{-1}, 10^{-1}\}$; $\{1,10\}$ \\\cline{7-8}
   S2.A    &  $3\cdot10^{-3}$      & 0                                &  $10^{-3}$        &   $2\cdot10^{-3}$    & $10^{-4}$                           & $c=10$: $\{0.1, 1, 4, 25\}$; $c=10^{-6}$: $\{0.1, 1, 4, 25\}$$\times10^{-7/4}$;  &$\{1,1,1,1\}$\\
   S2.B    &  $3\cdot10^{-3}$      & $9\cdot10^{-4}$         &  $10^{-3}$        &   $2\cdot10^{-3}$    & $10^{-4}$                           &$c=10^{-8}$: $\{0.1, 1, 4, 25\}$$\times10^{-9/4}$ &\\\cline{7-8}
   S3.A    &  $2\cdot10^{-4}$      & 0                                &  $10^{-3}$        &   $3\cdot10^{-2}$    & $10^{-4}$                            &$c=10^{-2}$: $\{0.01, 0.1, 1, 10\}$; $c=10^{-4}$:  &$\{1,1,1,1\}$\\          
   S3.B    &  $2\cdot10^{-4}$      & $6\cdot10^{-5}$         &  $10^{-3}$        &   $3\cdot10^{-2}$    & $10^{-4}$                            &$\{0.1, 1, 10, 100\}$$\times10^{-1/2}$; $c=10^{-8}$: $\{0.1, 1, 10, 100\}$$\times10^{-3/2}$  &\\\cline{7-8}
   S4.A    &  $10^{-3}$                & $10^{-5}$                   &  $10^{-3}$        &   $6\cdot10^{-3}$    & $10^{-4}$                            &10 & $c=10$: 100; $c=20$: 200;\\ 
   S4.B    &  $10^{-3}$                & $0.3\cdot10^{-3}$      &  $10^{-3}$        &   $6\cdot10^{-3}$    & $10^{-4}$                            & &$c=100$: 1000\\ 
   
   \end{tabular}
   }
\caption{\label{tab:kinetic_parameter} Reaction rate constants used in the mesoscopic simulations of the gene regulation kinetics. Values in the curly brackets correspond to the effective rate constants of binding (or unbinding) to the consecutive binding sites.}
  \end{table*}

% % Create the reference section using BibTeX:
% \bibliography{basename of .bib file}

\bibliography{autoreg1d_arxiv}
\end{document}